\documentclass[iop]{emulateapj}
\usepackage{epstopdf}
\usepackage{float}
\usepackage{subfigure}
\usepackage{lineno}
\usepackage{mathtools}
\usepackage{array}
\usepackage{rotating}

\usepackage{graphicx, subfigure}
\usepackage{float}
\graphicspath{{plots/} }

\received{June 2015}
\revised{revision date}
\accepted{acceptance date}

\shorttitle{Non-elastic processes in atom - Rydberg atom collisions}

\shortauthors{Mihajlov et al.}

\begin{document}
\title{Non-elastic processes in atom - Rydberg atom collisions: Review of state of art and problems}

\author{A.A.Mihajlov, V.A.Sre\'ckovi\'c, Lj. M. Ignjatovi\'c}
\affil{Institute of physics,Univesity of Belgrade,  P.O. Box 57, 11001, Belgrade, Serbia}

\email{vlada,mihajlov@ipb.ac.rs}

\author{A. N. Klyucharev}
\affil{Dept. of Physics, St-Petersburg University, Ulianovskaya 1, 198904 St. Petersburg, Petrodvorets, Russia}
\email{anklucharev@gmail.com}

\author{M.S.Dimitrijevi\'c}
\affil{Astronomical Observatory, Volgina 7, 11060 Belgrade 74,
       Serbia and Observatoire de Paris, 92195 Meudon Cedex, France \\and IHIS Techno experts, Batajni\v cki put 23, 11080 Zemun, Serbia}
\email{mdimitrijevic@aob.rs}

\and

\author{N.M. Sakan}
\affil{Institute of physics,Univesity of Belgrade,  P.O. Box 57, 11001, Belgrade, Serbia}

\email{nsakan@ipb.ac.rs}

\begin{abstract}
In our previous research, it has been demonstrated that such inelastic processes in atom Rydberg-atom collisions,
as chemi-ionization and (n-n') mixing, should be considered together. Here we will review the present state of the art
and the actual problems will be discussed. In this context, we will consider the influence of the
(n-n')-mixing during a symmetric  atom Rydberg-atom collision processes on the intensity of chemi-ionization process.
It will be taken into account H(1s) + H*(n) collisional systems, where the principal quantum number n $>>$ 1.
It will be demonstrated that the inclusion of (n-n') mixing in the calculation,  influences  significantly on the
values of chemi-ionization rate coefficients, particularly in the lower part of the block of the Rydberg states.
Different possible channels of the (n-n')-mixing influence on chemi-ionization  rate coefficients will be demonstrated.
The possibility of interpretation of the (n-n')-mixing influence  will be considered on the basis of two existing
methods for describing of the inelastic processes in  symmetrical atom Rydberg-atom collisions.
\end{abstract}


\section{Introduction}

Exploring and improving the new calculation possibilities and simulation techniques,
attracted extensive attention in the chemi-ionization and (n-n')-mixing processes
in atom –Rydberg atom collisions, which resulted in numerous papers dedicated to this
problem in various research fields like astrophysics, plasma physics, chemistry (see for example
\citet{boh12,bar07,mih07a,rya05}).

Two groups of inelastic processes in slow atom-Rydberg atom collisions will be considered in this paper: the chemi-ionization processes,
\begin{subequations}
\begin{align}
 A^{*}(n,l)+A \rightarrow A + A^{+} + \vec{e}
\label{eq:hemi1}\\
 A^{*}(n,l)+A \rightarrow A_{2}^{+} + \vec{e}
 \label{eq:hemi2}
\end{align}
\end{subequations}
and the processes of (n-n')-mixing
\begin{equation}
\label{eq:mix}
A^{*}(n,l)+A \rightarrow A + A^{*}(n',l').
\end{equation}
Here $A$ and $A^{*}(n,l)$ denote atom in the ground and in highly
excited (Rydberg) state with the given principal and orbital quantum
numbers $n$ and $l$, $A^{+}$ and $\vec{e}$ - atomic ion in the
ground state and free electron, while $A_{2}^{+}$ denotes the
molecular ion in the ground state.

\begin{figure}[h!]
\centerline{\includegraphics[width=\columnwidth,
height=0.75\columnwidth]{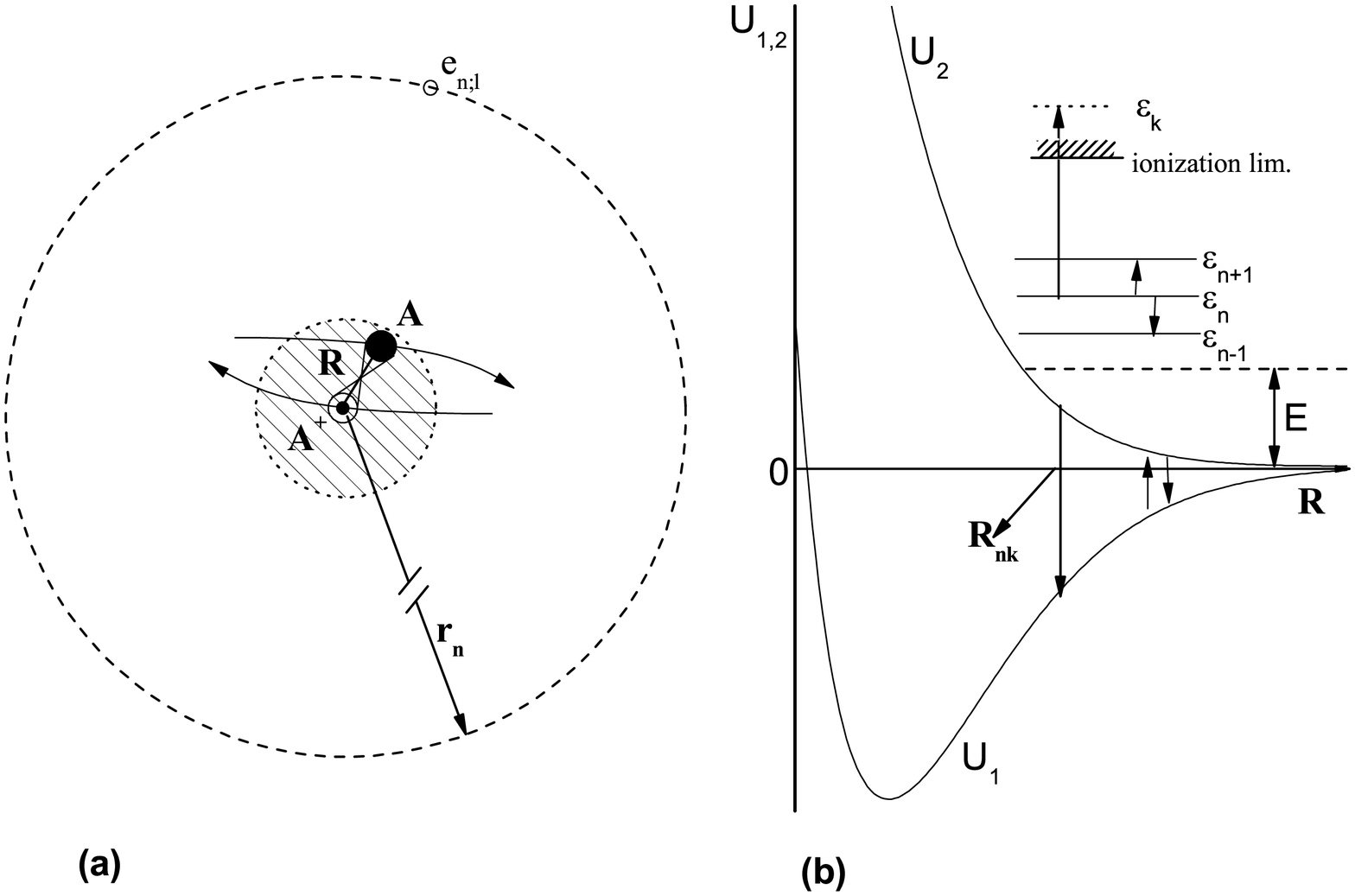}} \caption{(a) Schematic
illustration of $A^{*}(n,l)+A$ collision within the domain of
internuclear distances $R \ll r_{n;l}$, where $r_{n;l}\sim n^{2}$ is
the characteristic radius of Rydberg atom $A^{*}(n,l)$; (b)
Schematic illustration of the simultaneous resonant transitions of
the outer electron from the initial bound to the final state and the
sub-system $A^{+}+A$ from initial excited to the final ground
electronic state. If the outer electron becomes free
$\varepsilon_{k} > 0$ the processes (1) occur, while if the outer
electron remains in the bound state $\varepsilon_{n'} < 0$ the
processes \ref{eq:mix} occur.} \label{fig:SHEMA1}
\end{figure}

The processes (1) and (2), illustrated by Figs.~\ref{fig:SHEMA1}a
and \ref{fig:SHEMA1}b, were examined and discussed in the literature
for a long time (see e.g.~\cite{mih81,jan87}). These processes are
conditioned by the dipole resonant mechanism which was described in
details in \cite{mih12}. Significant contribution of processes (1)
and (2) in modeling of solar atmosphere is shown in
\citet{mih11,mih11b,bar07,mas09,mas10}, while the papers of
\citet{mih03} and \citet{sre13}, are devoted to the influence of these processes on
the kinetic of helium-rich star atmospheres.
Another important thing is that the presented results suggest that these processes,
due to their influence on free electron density and excited state populations in the atmospheres of M red dwarfs,
should also influence the atomic spectral line shapes (see e.g. \citet{mih07b}).

In spite of the fact that the processes (1) and (2) are caused by
the same mechanism, they are considered separately up to now. The
main aim of this work is to determine the influence of processes (2)
on the processes of chemi-ionization (1a) and (1b). Namely, already
from Fig. 1b one can notice the following: in the case when the
considered atomic collision proceed in accordance with the excited
molecular term  $U_{2}(R)$, before it enters in the zone where the
chemi-ionization processes (1a,b) occur, the system $A^{*}(n,l)+A$
passes through the zone where the processes (2) with $n'>n$ take
place.

As first, the way of inclusion of process (2) in the procedure of
calculation of rate coefficients of the chemi-ionization processes
(1a,b) will be described. For this purpose their values will be
determined under the conditions characteristic for the Solar
photosphere in the case $A=\textrm{H}(1s)$ and compared with the
rate coefficients of the same chemi-ionization processes, determined
in \citet{mih11}, but without inclusion of (n-n')-mixing processes.
We draw attention that, as a difference from this previous article,
chemi-ionization rate coefficients are here without the
simplification of the expression for Gaunt factor, connected with
the photo-ionization cross sections for the transitions of Rydberg
electron $\varepsilon(n,l)\rightarrow\varepsilon(k)$. Besides, here,
as a difference from \citet{mih11}, the average chemi-ionization
rate coefficient for a given $n$  is obtained as a result of the
corresponding averaging of partial chemi-ionization rate
coefficients for every $l$ where $0\leq l \leq n-1$.

Atomic units will be used throughout the paper.

\section{THE THEORY}

\subsection{General formulas}

Let $K_{1a}(n,l;T)$ and $K_{1b}(n,l;T)$ are rate coefficients of processes (1a) and (1b), separately determined for given $n$, $l$ and $T$ where $T$ is temperature of the considered plasma, and $K_{1}(n,l;T)$ is the total rate coefficient of processes (1a) and (1b) together, namely $K_{1}(n,l;T)=K_{1a}(n,l;T)+K_{1b}(n,l;T)$.

Because of further applications, we will then determine the average total rate coefficient
\begin{equation}
\label{eq:rate}
K_{1;n}(T)=\frac{1}{n^{2}}\cdot\sum_{l=0}^{n-1}(2l+1)\cdot K_{1}(n,l;T),
\end{equation}

\noindent and average rate coefficient of associative ionization $K_{1b;n}(T)$

\begin{equation}
\label{eq:rate_as}
K_{1b;n}(T)=\frac{1}{n^{2}}\cdot\sum_{l=0}^{n-1}(2l+1)\cdot K_{1b}(n,l;T).
\end{equation}

Partial rate coefficients $K_{1}(n,l;T)$ and $K_{1b}(n,l;T)$ are determined on the basis of standard expressions

\begin{equation}
\label{eq:K1}
K_{1}(n,l;T)=\int_{E_{n;i}}^{\infty}\sigma_{1}(n,l;E)\left(\frac{2E}{\mu_{red}}\right)^{1/2}f_{T}(E)dE,
\end{equation}
\begin{equation}
\label{eq:K1b}
K_{1b}(n,l;T)=\int_{E_{n;i}}^{\infty}\sigma_{1b}(n,l;E)\left(\frac{2E}{\mu_{red}}\right)^{1/2}f_{T}(E)dE,
\end{equation}

\noindent where $E$ is impact energy, $\sigma_{1}(n,l;E)$ and $\sigma_{1b}(n,l;E)$  are the corresponding cross sections, $\mu_{red}$ is reduced mass of the subsystem H(1s)+H$^{+}$, and $f_{T}(E)$ is the Maxwell distribution function: $f_{T}(E)=\exp(-E/kT)\sqrt E$. Parameter $E_{n;i}$ is given here with the relation $E_{n;i}=U_{2}(R_{n;i})$
where $R_{n;i}$ is the upper limit of the chemi-ionization zone which is the root of the equation $U_{12}=1/2n^{2}$.

The mentioned cross sections are determined here within the semi-classical approximation, with the help of also standard expressions
\begin{equation}
\label{eq:sig1}
\begin{split}
\sigma_{1}(n,l;E)=2\pi \int_{0}^{\rho_{1;max}}P_{1}(n, l;\rho; E)\rho d\rho,\\ \sigma_{1b}(n,l;E)=2\pi \int_{0}^{\rho_{1b;max}}P_{1b}(n, l;\rho; E)\rho d\rho
\end{split}
\end{equation}

\noindent where $\rho$ is impact parameter, $\rho_{1;max}$ and $\rho_{1b;max}$ - are the corresponding maximal values of this parameter, and $P_{1}(n, l;\rho; E)$ and $P_{1b}(n, l;\rho; E)$ - are the total probability of chemi-ionization and the probability of associative ionization, respectively determined for the given values of $n$, $l$, $\rho$ and $E$. These probabilities we will determine here in the form

\begin{equation}
\label{eq:P1}
P_{1}(n, l;\rho; E)=\frac{1}{2}\cdot p_{keep}(n, l;\rho; E)\cdot p_{i;1}(n, l;\rho; E),
\end{equation}
\begin{equation}
\label{eq:P1b}
P_{1b}(n, l;\rho; E)=\frac{1}{2}\cdot p_{keep}(n, l;\rho; E)\cdot p_{i;1b}(n, l;\rho; E)
\end{equation}

\noindent where $1/2$ is the probability that the subsystem  H(1s)+H$^{+}$ develops in accordance with the term $U_{2}(R)$, $ p_{keep}(n, l;\rho; E)$- probability that in the domain of values of $R$ where the processes (2) with $n'>n$ are possible, the state of this subsystem is held on, i.e. the excited electronic state with the energy $U_{2}(R)$, while $p_{i;1}(n, l;\rho; E)$  and $p_{i;1b}(n, l;\rho; E)$ are the corresponding ionization probabilities determined under the condition that subsystem  H(1s)+H$^{+}$ enters in the ionization zone with probability equal to 1.

\subsection{Probability  of ionization decay}
\label{sec:tech}

Similarly as in the previous papers, probabilities $p_{i;1}(n,l;\rho;E)$  and $p_{i;1b}(n,l;\rho;E)$ are determined here within the quasi-static decay approximation. Since these probabilities are determined in the similar way as in previous works of  \citet{mih07a,mih11}, here they are taken in the form

\begin{equation}
\begin{split}
\label{eq:pi1,b}
p_{i;1}(n, l;\rho; E)=1.0-\exp(-2q_{i;1}),\\
p_{i;1b}(n, l;\rho; E)=\exp(-q_{i;2})\cdot[1.0-\exp(-2q_{i;as})],
\end{split}
\end{equation}
\noindent where the quantities $q_{i;1}$, $q_{i;2}$ and $q_{i;as}$ are given as

\begin{equation}
\begin{split}
\label{eq:q1,2}
q_{i;as}=q_{i;1}-q_{i;2}, \quad q_{i;1}=\int_{R_{0}}^{R_{n;i}}\frac{W_{i}(n,l;R)}{\upsilon_{rad}(E,\rho;R)}dR,\\
q_{i;2}=\int_{R_{1b;max}}^{R_{n;i}}\frac{W_{i}(n,l;R)}{\upsilon_{rad}(E,\rho;R)}dR,
\end{split}
\end{equation}

The rate coefficient of ionization decay $W_{i}(n,l;R)$  and radial ion-atom velocity $\upsilon_{rad}(E,\rho;R)$ are given by expressions
\begin{equation}
\begin{split}
\label{eq:w1v}
W_{i}(n,l;R)=\frac{1}{ 2\pi}\cdot c\cdot U_{12}^{3}(R)\cdot D_{12}^{2}(R)\cdot\sigma_{ph.i}(n,l,\varepsilon_{ph}),\\
\upsilon_{rad}(E,\rho;R)=\left( \frac{2}{\mu_{red}}\left[ E-U_{2}(R)-\frac{E\rho^{2}}{R^{2}}\right] \right)^{1/2},
\end{split}
\end{equation}
\noindent where $c$ is the speed of light, $D_{12}=|<1|\hat{d}_{m.i}|2>|$ - molecular-ion dipole matrix element, $\sigma_{ph.i}(n,l,\varepsilon_{ph})$ - cross section for photoionization of excited hydrogen atom H$^{*}(n,l)$ by a photon with energy $\varepsilon_{ph}=U_{12}(R)$, and $U_{12}(R)=U_{2}(R)-U_{1}(R)$.

In the expression for dipole matrix element  $\hat{d}$  denotes operator of ion dipole momentum $H_{2}^{+}$ and $|1>$ and $|2>$  -  ground and first excited state of this ion.

In Eq. (\ref{eq:q1,2}) with $R_{0}$ is denoted the lower limit of the domain $R$ which is reached  during the collision with a given $\rho$ and $E$ , and with $R_{1b;min}$- the upper limit of the domain $R$ where is possible only the process of associative ionization (\ref{eq:hemi2}). Consequently, parameters $R_{0}$ represent here the roots of the equation: $U_{2}(R)=E\cdot(1-\rho^{2}/R^{2})$ , and $R_{1b;max}$ -root of the equation: $U_{12}(R)=E$ . Let draw attention that it is assumed in expressions (\ref{eq:pi1,b}) and (\ref{eq:q1,2}) that $R_{1b;max}<R_{n;i}$  In the case of $R_{1b;max}>R_{n;i}$ we have that the quantity $q_{i;2}=0$ , and $q_{i;as}=q_{i;1}$.

We draw attention that already at this point exist a difference compared to previous works concerning the chemi-ionization processes in stellar atmospheres \citep{mih07a,mih11}. Namely, in just mentioned works, the chemi-ionization rate coefficients were determined with the averaged ionization decay rate, obtained by averaging of partial rates over the whole shell with given $n$. This gives possibility to use the average over shell Kramers photo-ionization cross-section  adjusted with the help of approximate Gaunt factor. As a difference, the rate coefficients $K_{1}(n,l;T)$ and $K_{1b}(n,l;T)$  were determined here on the basis of Eqs. (\ref{eq:K1}) and (\ref{eq:K1b}) with the help of partial cross sections for photo-ionization, determined here on the basis of exact expressions from \citet{sob79}.

\subsection{Probability of pre-ionization decay}

From Eqs. (\ref{eq:P1}) and (\ref{eq:P1b}) one can notice that the basic difference, in comparison with previous papers, represents direct taking into account of the effect of decay of the initial electronic state of the considered atom-Rydberg atom system, due to the possibility of execution of excitation processes (\ref{eq:mix}) with $n'>n$. This one takes into account by the introduction of probability of maintenance of this state $p_{keep}(n,l;\rho;E)$. One determines this probability on the basis of the modified version of approximate method described in \citet{mih04} dedicated to the  $(n-n')$-mixing processes. Let remind, that the essence of this method is that at given $n$  each block  of Rydberg states from $n'=n+p_{1}$ to $n'=n+p_{2}$ is "spreading" in a part of "quasicontinuum" limited by values $n+p_{1}-\delta_{n}$  and $n+p_{2}+1-\delta_{n}$ , where the parameters  $\delta_{n}$ are determined  from the condition of maintainance of total number of states and total oscillator strengths for transitions from initial state of Rydberg electron to all states of  the separated block.  The mentioned modification has been conditioned with the fact that in the just mentioned work was determined an average rate of decay of the initial state  of system  connected with the transition of Rydberg electron from the state with the given $n$  in  states with $n'=n+p$ , where $p\ge 1$, while here we must to consider transitions of Rydberg electron  from the state $|n,l>$ to the states $|n+p,l-1>$ and $|n+p,l+1>$. In accordance with the just said, it is considered here that the preionization zone form the domain of internuclear distances such that $R_{n;i}<R<R_{n;n+1-\delta_{n}}$ , where $\delta_{n}=0.5\cdot[1-(1/3)\cdot O(1/n)]$, and domains $R$ corresponding to the mentioned transitions  with $p=1,2,3...$  make intervals $(R_{n;n+2-\delta_{n}},R_{n;n+1-\delta_{n}})$, $(R_{n;n+3-\delta_{n}},R_{n;n+2-\delta_{n}})$, $(R_{n;n+4-\delta_{n}},R_{n;n+3-\delta_{n}})$  … . The limits of these domains   $R_{n;n+p-\delta_{n}}$ are roots of the equations: $U_{12}(R)=0.5\cdot[1/n^{2}-1/(n+p-\delta_{n})^{2}]$.

In this work are taken into account the transitions with $1\le p\le 5$. Consequently, the probability $p_{keep}(n,l;\rho;E)$ could be represented as
\begin{equation}
\label{eq:p_k}
p_{keep}(n,l;\rho;E)=\prod_{p=1}^{5}p_{p;keep}(n,l;\rho;E),
\end{equation}
where $p_{p;keep}(n,l;\rho;E)$  is the probability of the maintenance of the initial state of the system within the interval  $(R_{n;n+p+1-\delta_{n}},R_{n;n+p-\delta_{n}})$.

Since the mechanism of the pre-ionization decay is the same as in the case of the ionization one, we take immediately that probabilities $p_{p;keep}(n,l;\rho;E)$ are given with the relations
\begin{equation}
\label{eq:p_k1}
\begin{split}
p_{p;keep}(n,l;\rho;E)=\exp(-x_{p}),\\
x_{p}=\int_{R_{n;n+p+1-\delta_{n}}}^{R_{p}}\frac{w_{n;n+p}(n,l:R)}{\upsilon_{rad}(E,\rho,R)},
\end{split}
\end{equation}
where the decay rate $w_{n;n+p}(n,l;R)$ is conditioned by the dipole
mechanism within the interval
$(R_{n;n+p+1-\delta_{n}},R_{n;n+p-\delta_{n}})$.

The upper limit $R_{p}$ is given by
\begin{equation}
R_{p}=
     \left\{
             \begin{array}{ll}
              \displaystyle{ R_{n;n+p-\delta_{n}}, R_{n;n+p-\delta_{n}} \le
R_{up;mix}(E,\rho)} \\
               \displaystyle{R_{up;mix}(E,\rho), R_{n;n+p-\delta_{n}}>R_{up;mix}(E,\rho)\ge
               R_{n;n+p}}
             \end{array}
     \right.
\label{eq:Rp}
\end{equation}
where $R_{n;n+p}$ is the resonant distance of the process
(\ref{eq:mix}) for given $n$ i $n'=n+p$ , determined as a root of
the equation
\begin{equation}
U_{12}(R)=\frac{1}{2}\cdot\left[\frac{1}{n^{2}}-\frac{1}{(n+p)^{2}}\right].
\label{eq:root}
\end{equation}
The parameter $R_{up;mix}(E,\rho)$ is separately discussed in
Appendix 1. Lets draw attention that here in the case when
$R_{up;mix}(E,\rho)<R_{n;n+p}$ is considered that
$p_{p;keep}(n,l;\rho;E)=0$ . In accordance with the above said, the
decay rate  $w_{n;n+p}(n,l;R)$ is given here by the relation
\begin{equation}
\label{eq:w_np}
\begin{split}
w_{n;n+p}(n,l;R)=\frac{2\pi}{3}\cdot U_{12}^{4}(R_{n;n+p})\cdot \widetilde{n}^{3}\cdot D_{12}^{2} \cdot r_{n,l;n+p}^{2},\\
\widetilde{n}=n\cdot[1-2n^{2}\cdot U_{12}(R)]^{-1/2},
\end{split}
\end{equation}
where $r_{n,l;n+p}^{2}=|<n,l|\hat{d}_{at}|n,l-1>|^{2} + |<n,l|\hat{d}_{at}|n,l+1>|^{2}$, $\hat{d}_{at}$  - is the operator of dipole moment of hydrogen atom, and  $|n,l>$, $|n,l-1>$  and $|n,l+1>$  denote the corresponding states of Rydberg electron.

\section{RESULTS AND DISCUSSION}

It follows from the above presented material that the total rate coefficients of the processes (\ref{eq:hemi1}) and (\ref{eq:hemi2}) together, and rate coefficients for the associative ionization (\ref{eq:hemi2}), i.e. $K_{1;n}(T)$ and $K_{1b;n}(T)$ are determined on the basis of  Eqs. (\ref{eq:rate})-(\ref{eq:w_np}). Lets draw attention that, strictly speaking, chemi-ionization processes (\ref{eq:hemi1}) and (\ref{eq:hemi2}) can be described  on the basis of dipole resonant mechanism only in the case of the state with $n \ge 5$, for which the potential curves of the system $\textrm{H}^{*}(n,l)+\textrm{H}(1s)$ lay above the potential curve of the system $\textrm{H}^{+}+\textrm{H}^{-}(1s^{2})$, where $\textrm{H}^{-}(1s^{2})$  is stable negative  hydrogen ion. However, it can be shown that the points of the intersection of potential curves of the system $\textrm{H}^{*}(n,l)+\textrm{H}(1s)$ with $n=2,3$ and $4$ with the potential curve of the system $\textrm{H}^{+}+\textrm{H}^{-}(1s^{2})$ are located on the internuclear distances, which are several times larger than the average atomic radius $\textrm{H}^{*}(n,l)$  so that the existence of these intersections can not significantly affect the values of the corresponding rate coefficients of the processes (\ref{eq:hemi1}) and (\ref{eq:hemi2}). Consequently the applicability of the dipole resonance mechanism for the states with $n<5$ depends to what degree     may be regarded as fulfilled condition $R_{n;n+1}\ll r_{n;l}$ where $r_{n;l}$ is the mean radius of the corresponding orbit of the outer electron. One notices that from this aspect, the dipole resonant mechanism can not be applied in the case of $n=2$, while in the case of the states $n=3$ and $4$ the application of this mechanism can be completely justified.

\begin{figure}
\centerline{\includegraphics[width=\columnwidth, height=0.8\columnwidth]{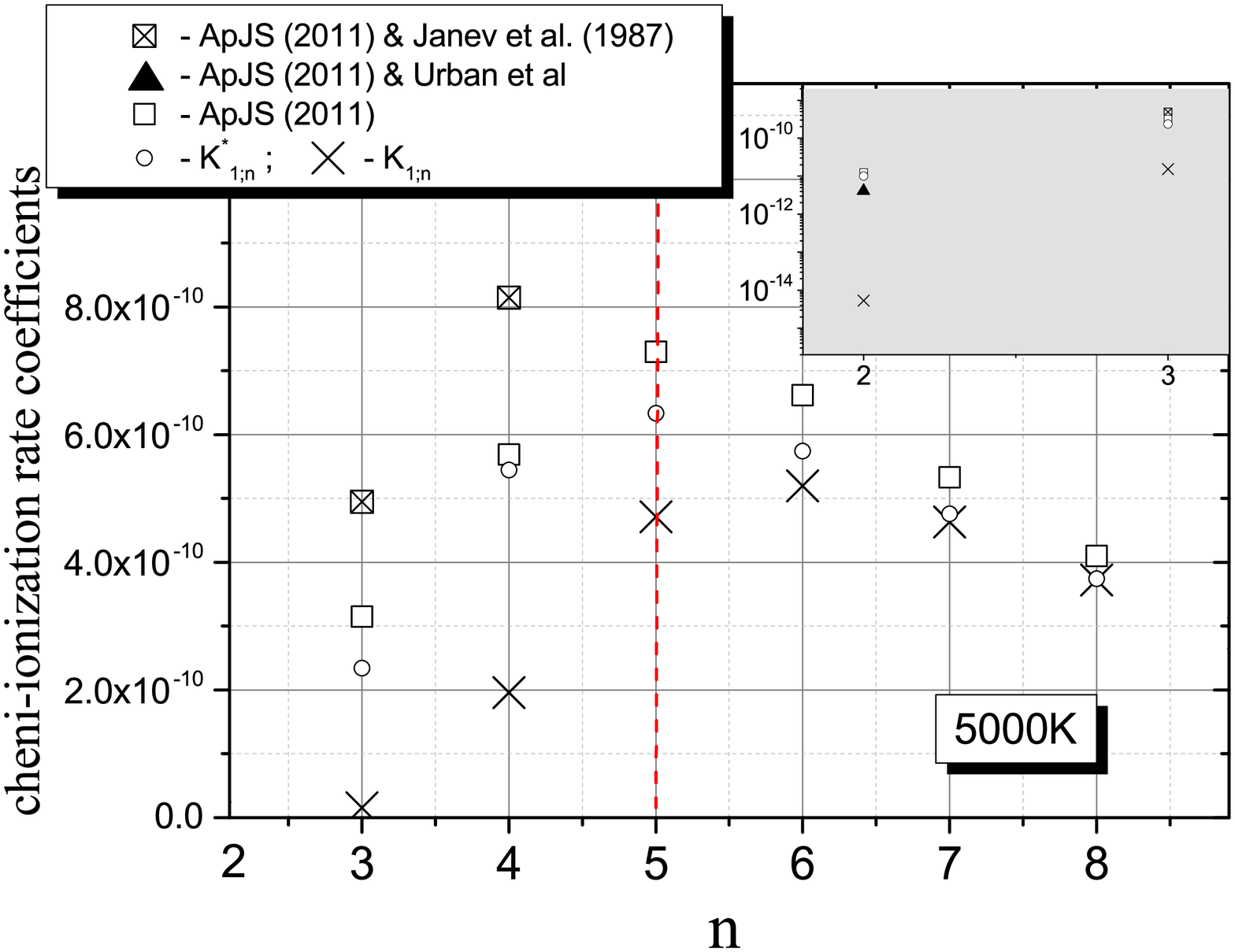}}
\caption{Comparison of the calculated values of rate coefficients of the chemi-ionization processes (\ref{eq:hemi1}) and (\ref{eq:hemi2}) with the data from \citet{mih11}.} \label{fig:2}
\end{figure}

Total values of the rate coefficients of chemi-ionization processes $K_{1;n}(T)$ within the range $3 \le n \le 15$ are presented in Tab. \ref{tab:Ksr}. Bearing in mind the main application, of here obtained  results, on the photosphere and lower chromosphere of the Sun, calculations of these rate coefficients were performed here for temperatures $4000 \textrm{K} \le T \le 10000 \textrm{K}$. The processes (\ref{eq:hemi2}) are characterized in this paper via the corresponding branch coefficient $X_{1b;n}(T)$ given as

\begin{equation}
\label{eq:X}
X_{1b;n}(T)=\frac{K_{1b;n}}{K_{1;n}}.
\end{equation}

Values of coefficients $X_{1b;n}(T)$ for the same $n$ and $T$ are presented in  Tab. \ref{tab:X}. In accordance with the above said, rate coefficients are determined here by summing the probability of the decay of the initial state of the collisional system in preionization  zone with Rydberg electron transitions from state $|n>$  to state $|n+p>$, where $1\le p \le 5$ .

In order to demonstrate significance of the presented calculation we will
compare the chemi-ionization rate coefficients $K_{i;n}(T)$ with the
corresponding rate coefficients $K_{i;n}^{*}(T)$ from \citet{mih11}.
Let us note that the coefficients $K_{i;n}^{*}(T)$ are obtained in the same way
as the coefficients $K_{i;n}(T)$ but taking $p_{keep}(n,l;\rho;E)=0$,
where $p_{keep}(n,l;\rho;E)$ is the total probability of the preionization decay
given by Eqs. (\ref{eq:p_k}) - (\ref{eq:w_np}).
All mentioned quantities are presented in Fig \ref{fig:2} for the case of $T=5000$ K. Lets draw attention that in relation to the previous work of \citet{mih11} in this figure are presented  not only the total rate coefficients, determined on the basis of dipole resonance mechanism for $3 \le n \le 8$ but also and rate coefficients  determined there on the basis of data from \citet{jan87} for $n=3$ and $4$, and from \citet{urb91} for $n=2$. One can notice from this figure that there are noticeable differences between the values of the rate coefficients  determined in \citet{mih11} and values $K_{i;n}^{*}(T)$, while the differences in relation to the rate coefficients $K_{i;n}(T)$  are very large  for $n \le 6$ and  decrease quickly  with the increase of $n$ in the area $n>6$.

In previous works \citep{mih03,mih07b} related to the photosphere of  a M red dwarf with temperature near to 4000 K, it has been shown that on populations of hydrogenic Rydberg states in this photosphere as well as on its other characteristics, influence strongly just the chemi-ionization processes (\ref{eq:hemi1}) and (\ref{eq:hemi2}) with $4 \le n \le 8$. It is clear that, already because of this, it is indispensable to take into account the changes  of rate coefficients of these processes, which, in accordance with our results, are particularly large for $n \le 6$. From the material presented here, follows also the  great importance of the further investigation of the properties of decay of the initial state of the collisional system $\textrm{H}^{*}(n,l)+\textrm{H}(1s)$  in the pre-ionization zone.

Additionally, obtained  here results suggest that the rate coefficients of the chemi-ionization  processes (\ref{eq:hemi1}) and (\ref{eq:hemi2}) could be affected and by other channels of influence of the processes (\ref{eq:mix}). Here we have in view the processes of (n-n ') mixing taking place in two or more steps.

\section{CONCLUSIONS}
\label{sec:conc}

In the presented work is shown that the processes of  (n-n')-mixing (\ref{eq:mix}) influence considerably on the rates of chemi-ionization processes  (\ref{eq:hemi1}) and (\ref{eq:hemi2}). Calculations, which characterize this influence on the quantitative level have been performed. As one can see from figure \ref{fig:2}, inclusion of the (n-n') mixing processes into consideration, reduce the chemi-ionization rate coefficients.
The obtained results are finalized in the tabular form, where the values of total constants for rates of the processes (\ref{eq:hemi1}) and (\ref{eq:hemi2}) together, and rates for the process of associative ionization (\ref{eq:hemi2}) are presented. The tables cover the range of values, of the principal quantum number of Rydberg states of Hydrogen atom, from $n=3$ to $n=15$ and the temperature range from $T=4000$ K to $T=10000$ K, so that they can be directly applied in connection with the modeling of photosphere and lower chromosphere of the Sun.  Moreover, in the work have been discussed further directions of the investigation of the influence of  (n-n')-mixing processes on the chemi-ionization processes taking into account (n-n') mixing processes which occur in two or more steps.

\acknowledgments

The authors are thankful to the Ministry of Education, Science and Technological Development
of the Republic of Serbia for the support of this work within the projects 176002, III44002 and 171014.


\newcommand{\noopsort}[1]{} \newcommand{\printfirst}[2]{#1}
  \newcommand{\singleletter}[1]{#1} \newcommand{\switchargs}[2]{#2#1}

\begin{sidewaystable}
\centering
\caption{Calculated values of the coefficient $K_{1;n}(T)$(cm$^{3}$ s$^{-1}$) as a function of $n$ and $T$.}
\begin{tabular}
{c c c c c c c c c c c c c c } \hline
   & \multicolumn{11}{c}{n}\\
   \cline{2-14}
  $T$ & 3 & 4 & 5 & 6 & 7 & 8 & 9 & 10 & 11 & 12 & 13 & 14 & 15\\
  \hline
4000  & 7.17E-12 & 1.54E-10 & 3.58E-10 & 4.28E-10 & 3.98E-10 & 3.30E-10 & 2.61E-10 & 2.06E-10 & 1.63E-10 & 1.28E-10 & 1.02E-10 & 8.14E-11 & 6.62E-11 \\
4250  & 9.01E-12 & 1.63E-10 & 3.88E-10 & 4.52E-10 & 4.15E-10 & 3.42E-10 & 2.69E-10 & 2.11E-10 & 1.66E-10 & 1.31E-10 & 1.04E-10 & 8.28E-11 & 6.72E-11 \\
4500  & 1.11E-11 & 1.72E-10 & 4.16E-10 & 4.76E-10 & 4.32E-10 & 3.53E-10 & 2.76E-10 & 2.16E-10 & 1.70E-10 & 1.33E-10 & 1.06E-10 & 8.40E-11 & 6.81E-11 \\
4750  & 1.33E-11 & 1.83E-10 & 4.43E-10 & 4.98E-10 & 4.48E-10 & 3.63E-10 & 2.83E-10 & 2.20E-10 & 1.73E-10 & 1.35E-10 & 1.07E-10 & 8.51E-11 & 6.89E-11 \\
5000  & 1.53E-11 & 1.96E-10 & 4.71E-10 & 5.20E-10 & 4.63E-10 & 3.72E-10 & 2.89E-10 & 2.24E-10 & 1.76E-10 & 1.38E-10 & 1.09E-10 & 8.62E-11 & 6.97E-11 \\
5250  & 1.73E-11 & 2.12E-10 & 4.98E-10 & 5.42E-10 & 4.77E-10 & 3.81E-10 & 2.95E-10 & 2.28E-10 & 1.78E-10 & 1.40E-10 & 1.10E-10 & 8.73E-11 & 7.05E-11 \\
5500  & 1.96E-11 & 2.31E-10 & 5.26E-10 & 5.63E-10 & 4.90E-10 & 3.89E-10 & 3.01E-10 & 2.31E-10 & 1.80E-10 & 1.41E-10 & 1.11E-10 & 8.84E-11 & 7.12E-11 \\
5750  & 2.30E-11 & 2.51E-10 & 5.53E-10 & 5.83E-10 & 5.03E-10 & 3.96E-10 & 3.06E-10 & 2.35E-10 & 1.82E-10 & 1.43E-10 & 1.13E-10 & 8.94E-11 & 7.19E-11 \\
6000  & 2.81E-11 & 2.71E-10 & 5.79E-10 & 6.03E-10 & 5.15E-10 & 4.04E-10 & 3.11E-10 & 2.38E-10 & 1.84E-10 & 1.44E-10 & 1.14E-10 & 9.03E-11 & 7.25E-11 \\
6250  & 3.53E-11 & 2.91E-10 & 6.03E-10 & 6.21E-10 & 5.26E-10 & 4.11E-10 & 3.16E-10 & 2.41E-10 & 1.86E-10 & 1.46E-10 & 1.15E-10 & 9.12E-11 & 7.30E-11 \\
6500  & 4.37E-11 & 3.11E-10 & 6.26E-10 & 6.39E-10 & 5.37E-10 & 4.17E-10 & 3.20E-10 & 2.44E-10 & 1.88E-10 & 1.47E-10 & 1.16E-10 & 9.19E-11 & 7.36E-11 \\
7000  & 6.01E-11 & 3.50E-10 & 6.70E-10 & 6.72E-10 & 5.59E-10 & 4.30E-10 & 3.28E-10 & 2.50E-10 & 1.92E-10 & 1.49E-10 & 1.18E-10 & 9.33E-11 & 7.46E-11 \\
7500  & 7.08E-11 & 3.90E-10 & 7.13E-10 & 7.03E-10 & 5.80E-10 & 4.43E-10 & 3.36E-10 & 2.55E-10 & 1.95E-10 & 1.51E-10 & 1.20E-10 & 9.46E-11 & 7.57E-11 \\
8000  & 7.91E-11 & 4.31E-10 & 7.54E-10 & 7.31E-10 & 5.99E-10 & 4.55E-10 & 3.44E-10 & 2.60E-10 & 1.98E-10 & 1.54E-10 & 1.21E-10 & 9.57E-11 & 7.68E-11 \\
8500  & 8.91E-11 & 4.71E-10 & 7.93E-10 & 7.57E-10 & 6.14E-10 & 4.65E-10 & 3.51E-10 & 2.64E-10 & 2.01E-10 & 1.56E-10 & 1.22E-10 & 9.66E-11 & 7.75E-11 \\
9000  & 9.91E-11 & 5.13E-10 & 8.27E-10 & 7.82E-10 & 6.27E-10 & 4.74E-10 & 3.56E-10 & 2.68E-10 & 2.04E-10 & 1.58E-10 & 1.23E-10 & 9.74E-11 & 7.81E-11 \\
9500  & 1.06E-10 & 5.56E-10 & 8.57E-10 & 8.06E-10 & 6.40E-10 & 4.82E-10 & 3.61E-10 & 2.72E-10 & 2.06E-10 & 1.59E-10 & 1.25E-10 & 9.82E-11 & 7.86E-11 \\
10000  & 1.07E-10 & 6.03E-10 & 8.82E-10 & 8.30E-10 & 6.55E-10 & 4.90E-10 & 3.66E-10 & 2.75E-10 & 2.08E-10 & 1.61E-10 & 1.26E-10 & 9.91E-11 & 7.93E-11 \\
 \hline
 \end{tabular}
\label{tab:Ksr}
\end{sidewaystable}

\begin{table}[htbp]
\begin{center}
\caption{Calculated values of the branch coefficient $X_{1b;n}$ as a function of $n$ and $T$.}
\label{tab:X}
\begin{tabular}
{c c c c c c c c c c c c c c } \hline
   & \multicolumn{12}{c}{n}\\
   \cline{2-14}
  $T$ & 3 & 4 & 5 & 6 & 7 & 8 & 9 & 10 & 11 & 12 & 13 & 14 & 15\\
  \hline
4000&0.684&0.608&0.458&0.365&0.306&0.243&0.218&0.208&0.201&0.186&0.169&0.154&0.137 \\
4250&0.607&0.563&0.437&0.346&0.284&0.232&0.211&0.202&0.195&0.178&0.160&0.143&0.129 \\
4500&0.543&0.519&0.421&0.329&0.265&0.222&0.206&0.198&0.189&0.170&0.152&0.132&0.122 \\
4750&0.497&0.475&0.407&0.314&0.248&0.213&0.201&0.194&0.184&0.162&0.144&0.122&0.114 \\
5000&0.467&0.431&0.395&0.301&0.232&0.205&0.197&0.190&0.180&0.155&0.137&0.112&0.108 \\
5250&0.473&0.427&0.370&0.284&0.223&0.201&0.193&0.185&0.172&0.146&0.130&0.109&0.104 \\
5500&0.467&0.419&0.347&0.269&0.215&0.197&0.189&0.181&0.164&0.137&0.123&0.107&0.101 \\
5750&0.442&0.411&0.328&0.254&0.208&0.193&0.186&0.177&0.157&0.128&0.116&0.104&0.098 \\
6000&0.397&0.403&0.310&0.242&0.201&0.190&0.183&0.173&0.150&0.119&0.110&0.101&0.095 \\
6250&0.349&0.380&0.294&0.229&0.197&0.187&0.179&0.166&0.142&0.116&0.107&0.099&0.092 \\
6500&0.308&0.360&0.279&0.218&0.194&0.184&0.175&0.159&0.134&0.113&0.104&0.096&0.089 \\
7000&0.263&0.327&0.254&0.198&0.187&0.178&0.169&0.146&0.118&0.106&0.098&0.092&0.083 \\
7500&0.223&0.292&0.234&0.190&0.180&0.172&0.157&0.133&0.112&0.101&0.093&0.086&0.078 \\
8000&0.199&0.263&0.216&0.183&0.173&0.167&0.146&0.120&0.105&0.096&0.087&0.081&0.073 \\
8500&0.198&0.243&0.194&0.178&0.169&0.161&0.134&0.112&0.100&0.092&0.083&0.075&0.070 \\
9000&0.198&0.225&0.175&0.173&0.165&0.156&0.123&0.105&0.095&0.087&0.080&0.069&0.067 \\
9500&0.201&0.210&0.164&0.167&0.161&0.142&0.116&0.099&0.091&0.084&0.075&0.067&0.065 \\
10000&0.218&0.196&0.155&0.161&0.156&0.129&0.109&0.095&0.087&0.081&0.071&0.066&0.063 \\

 \hline
 \end{tabular}
 \end{center}
\end{table}

\begin{table}[htbp]
\begin{center}
\caption{Calculated values of the parameters which characterize pre-ionization zone.
Phase $\varphi(R_{n,n+1},E_{n;i};\rho=0)$ is given by the relation Eq. (A2).} \label{tab:phi}
\begin{tabular}
{c c c c c c } \hline
  $n$ & $R_{ni}$&$E_{n;i}=U2(Rn;i)$&$R_{n,n+1}$&$\varphi(R_{n,n+1},E_{n;i};\rho=0)$&$P_{c.exc}(\varphi(R_{n,n+1},E_{n;i};\rho=0))$ \\
  \hline
3&4.79&0.02738&5.839&1.840&0.92929 \\
4&5.52&0.01431&6.777&1.143&0.82824 \\
5&6.08&0.00871&7.497&0.782&0.49618 \\
6&6.52&0.00581&8.087&0.567&0.28891 \\
7&6.89&0.00413&8.580&0.433&0.17619 \\
8&7.21&0.00306&9.010&0.341&0.11201 \\
9&7.49&0.00234&9.380&0.278&0.07544 \\
10&7.73&0.00183&9.725&0.229&0.05167 \\
11&7.95&0.00146&10.035&0.193&0.03667 \\
12&8.16&0.00119&10.317&0.165&0.02691 \\
13&8.34&0.00098&10.551&0.146&0.02108 \\
14&8.51&0.00081&10.839&0.122&0.01489 \\
15&8.66&0.00068&11.002&0.114&0.01297 \\
 \hline
 \end{tabular}
 \end{center}
\end{table}

\appendix

\section{Appendix material}

The characteristic length $R_{up;mix}$ is defined here as the upper limit of the domain $R$ where at given $E$ and $\rho$ we can consider that the inner electron is in the subsystem $\textrm{H}^{+}+\textrm{H}(1s)$ and that is sufficiently delocalized, so that this subsystem could be treated as a quasi-molecular complex.  As a qualitative characteristics of the mentioned delocalization, one takes here the probability of resonant charge exchange $P_{c.exc}(R;E;\rho)$  in the subsystem  $\textrm{H}^{+}+\textrm{H}(1s)$  as a function of  $R$ at given $\rho$ and $E$. As the basis for this, is taken the theory of the process: $\textrm{H}^{+}+\textrm{H}(1s)\rightarrow \textrm{H}(1s) + \textrm{H}^{+}$, developed in \citet{fir51} and \citet{bat62}. From this theory follows that
\begin{equation}
\label{eq:A1}
P_{c.exc}(R;E;\rho)=\sin^{2}(\varphi(R;E;\rho)),
\end{equation}
where the phase $\varphi(R;E;\rho)$ is given by the relation
\begin{equation}
\label{eq:A2}
\varphi(R;E;\rho)=\frac{1}{2}\int_{R}^{\infty}\frac{U_{12}(R')}{\upsilon_{rad}(R',\rho,E)}dR',
\end{equation}
which can be used in the considered case since $P_{c.exc}(R;E;\rho)$  becomes noticeably different from zero only deeply inside the orbit of Rydberg electron at given  $n$. On the basis of data from   \citet{fir51} and \citet{bat62}  can be considered that in the case when  $P_{c.exc}(R;E;\rho)$  reaches the value of  $1/2\pi$ the corresponding  $R$  may be considered as the upper limit of the charge exchange zone at given  $\rho$ and $E$. And consequently, as the upper limit of domain with a sufficient degree of delocalization of electron in the subsystem $\textrm{H}^{+}+\textrm{H}(1s)$. Consequently, the parameter $R_{up;mix}$ is determined here as the root of equation
\begin{equation}
\label{eq:A3}
\sin^{2}(\varphi(R;E;\rho))=\frac{1}{2\pi}
\end{equation}
where $\varphi(R;E;\rho)$  is given by  Eq. (\ref{eq:A2}) under condition that this root is in the domain of monotonical increase of the left side of  Eq. (\ref{eq:A3}).

The behavior of phase $\varphi(R;E;\rho)$  is illustrated by  Tab.~\ref{tab:phi}, where its values for  $E=E_{n;i}$, $\rho=0$ and $R=R_{n;n+1}$ within the range $3 \le n \le 15$ are shown. Of course, these data should be treated as the qualitative ones  since Eqs. (\ref{eq:A1}) and (\ref{eq:A2}) have strict sense in the case $E \gg U_{12}(R)$ while in our case when this condition is fulfilled only for $n>7$.

\end{document}